\documentclass [floatfix,twocolumn,amssymb,aps,prl,showpacs,superscriptaddress]{revtex4-1}

\usepackage{graphicx}
\usepackage{float}
\usepackage{bm}
\usepackage{xcolor}
\usepackage{silence}

\begin{document}

\title{Competition between orthorhombic and re-entrant tetragonal phases in\\ underdoped Ba$_{1-x}$K$_x$Fe$_2$As$_2$ probed by the response to controlled disorder }
\author{E.~I.~Timmons}
\affiliation{Ames Laboratory US DOE, Ames, Iowa 50011, USA}
\affiliation{Department of Physics and Astronomy, Iowa State University, Ames, Iowa 50011,
USA }

\author{M.~A.~Tanatar}
\email{tanatar@ameslab.gov}
\affiliation{Ames Laboratory US DOE, Ames, Iowa 50011, USA}
\affiliation{Department of Physics and Astronomy, Iowa State University, Ames, Iowa 50011,
USA }

\author{K.~Willa}
\affiliation{Materials Science Division, Argonne National Laboratory, 9700 S. Cass Avenue, Argonne, Illinois 60439, USA}

\author{S. Teknowijoyo}
\affiliation{Ames Laboratory US DOE, Ames, Iowa 50011, USA}
\affiliation{Department of Physics and Astronomy, Iowa State University, Ames, Iowa 50011,
USA }

\author{Kyuil Cho}
\affiliation{Ames Laboratory US DOE, Ames, Iowa 50011, USA}

\author{M. Ko\'nczykowski}
\affiliation{Laboratoire des Solides Irradi\'es, \'Ecole Polytechnique, CNRS, CEA, Universit\'e Paris-Saclay, 91128 - Palaiseau Cedex, France}

\author{O.~Cavani}
\affiliation{Laboratoire des Solides Irradi\'es, \'Ecole Polytechnique, CNRS, CEA, Universit\'e Paris-Saclay, 91128 - Palaiseau Cedex, France}

\author{Yong~Liu}
\affiliation{Ames Laboratory US DOE, Ames, Iowa 50011, USA}

\author{T.~A.~Lograsso}
\affiliation{Ames Laboratory US DOE, Ames, Iowa 50011, USA}
\affiliation{Department of Materials Science and Engineering, Iowa State University, Ames, Iowa 50011, USA }

\author{U.~Welp}
\affiliation{Materials Science Division, Argonne National Laboratory, 9700 S. Cass Avenue, Argonne, Illinois 60439, USA}

\author{R.~Prozorov}
\email{prozorov@ameslab.gov}
\affiliation{Ames Laboratory US DOE, Ames, Iowa 50011, USA}
\affiliation{Department of Physics and Astronomy, Iowa State University, Ames, Iowa 50011,
USA }

\date{19 October 2018}

\begin{abstract}

Low-temperature (22~K) irradiation with 2.5~MeV electrons was used to study the competition between stripe ${\rm C_2}$ and tetragonal ${\rm C_4}$ antiferromagnetic phases which exist in a narrow doping range around $x=$0.25 in hole-doped Ba$_{1-x}$K$_x$Fe$_2$As$_2$. In nearby compositions outside of this range, at $x=$0.22 and $x=$0.19, the temperatures of both the concomitant orthorhombic/stripe antiferromagnetic transition $T_{\rm C2}$ and the superconducting transition $T_{\rm c}$ are monotonically suppressed by added disorder at similar rates of about 0.1~K/$\mu \Omega$cm, as revealed through using resistivity variation as an intrinsic measure of scattering rate. In a stark contrast, a rapid suppression of the ${\rm C_4}$ phase at the rate of 0.24 K/$\mu \Omega$cm is found at $x=$0.25. Moreover, this suppression of the ${\rm C_4}$ phase is accompanied by unusual disorder-induced stabilization of the ${\rm C_2}$ phase, determined by resistivity and specific heat measurements. The rate of the ${\rm C_4}$ phase suppression is notably higher than the suppression rate of the spin-vortex phase in the Ni-doped CaKFe$_4$As$_4$ (0.16 K/$\mu \Omega$cm).

\end{abstract}



\maketitle


Cooper pair binding mediated by magnetic fluctuations \cite{Mathur} is actively discussed as a possible mechanism of superconductivity in several classes of unconventional superconductors including heavy fermions \cite{heavyfermionLonzarich}, high-$T_{\rm c}$ cuprates \cite{LouisNatureQCP} and, more recently, iron-based superconductors \cite{Paglione}. A fingerprint of this model is the observation of the highest superconducting transition temperature, $T_{\rm c}$, coinciding with a quantum critical point (QCP) where the temperature of the magnetic transition, $T_N$, goes to zero at a point in a $T-x$ phase diagram with $x$ being a non-thermal control parameter such as doping, pressure, magnetic field or disorder \cite{Mathur,review,KchoNbSe2}. Strong magnetic fluctuations at the QCP lead to non-Fermi liquid behavior of all electronic properties, for example logarithmic divergence of the heat capacity and $T$-linear electrical resistivity.

In iron-based superconductors, this phenomenology is clearly observed in isovalent P-substituted BaFe$_2$(As$_{1-x}$P$_x$)$_2$ (Ba122) \cite{PdopedQCP,Pdopedcaxis,PAnalytis},  however it fails in hole-doped Ba$_{1-x}A_x$Fe$_2$As$_2$ ($A$=Na,K) compositions which have the highest $T_{\rm c}$. Here, the suppression of the transition temperature $T_{\rm C2}$ of the orthorhombic antiferromagnetic phase with stripe pattern of in-plane moments (${\rm C_2}$ phase) \cite{stripe,stripe2} does not proceed monotonically to zero, but is interrupted by the emergence of a new tetragonal  ${\rm C_4}$ magnetic phase below temperature $T_{\rm C4}$ \cite{Elena,Elena2,Anna,AvciNa,AvciK,entropy}. Being in very close proximity to the highest $T_{\rm c}$ doping range, this phase may play an important, yet not understood, role in the superconducting pairing \cite{Furukawa}.

The ${\rm C_4}$ phase is also observed in other hole-doped 122-type compounds, Ca$_{1-x}$Na$_x$Fe$_2$As$_2$ \cite{CaNa}, Sr$_{1-x}$Na$_x$Fe$_2$As$_2$ \cite{SrNa} and Ba(Fe$_{1-x}$Mn$_x$)$_2$As$_2$ \cite{Mn}. The ${\rm C_4}$ phase in Sr$_{1-x}$Na$_x$Fe$_2$As$_2$ was shown to be a double-$Q$ spin-charge density wave, with a moment equal to zero on every second iron atom \cite{doubleQ}. A similar ${\rm C_4}$ phase but with a different type of magnetic order is found in electron-doped CaK(Fe$_{1-x}$$TM_x$)$_2$As$_2$, with $TM$=Co, Ni  \cite{hedgehog}.  Theoretically, the origin of this phase has been attributed to itinerant magnetism \cite{RMF,Efremov}, magnetic moments with effects of frustration \cite{QSi} or the effects of spin-orbit coupling \cite{Orth1,Orth2}.

It was recently suggested, that disorder can lead to a stabilization of the spin-charge density wave ${\rm C_4}$ phase as compared to the ${\rm C_4}$ spin vortex state and the ${\rm C_2}$ phase in the phase diagram of the hole-doped compositions \cite{disorderSchmalian}. Motivated by this theoretical prediction, we report here a study on the effect of electron irradiation in hole-doped Ba$_{1-x}$K$_x$Fe$_2$As$_2$ with $x=$0.25 revealing clear signatures of the ${\rm C_4}$ phase in temperature-dependent resistivity and heat capacity measurements. For reference, we also study the effect of electron irradiation on nearby compositions with $x=$0.19 and $x=$0.22 outside the ${\rm C_4}$ phase doping range. We find that disorder suppresses the ${\rm C_4}$ phase at a rate which is significantly higher than the suppression rate of the ${\rm C_2}$ phase in nearby compositions and in the spin-vortex phase of CaK(Fe$_{1-x}$Ni$_x$)$_4$As$_4$ \cite{Serafim}. It also leads to an unusual slight increase of $T_{\rm C2}$ suggesting its stabilization with disorder. Our results clearly show competition between these two types of magnetic orders.


Single crystals of Ba$_{1-x}$K$_x$Fe$_2$As$_2$ were grown as described in detail in Ref.~\onlinecite{YLiucrystals}. Large, above 5$\times$5 mm$^2$ surface area crystals were cleaved on both sides to a thickness of typically 0.1 mm to minimize the variation of the K-content with thickness. The cleaved slabs were  characterized by electron-probe microanalysis with wavelength dispersive spectroscopy (WDS). The crystals from three different batches were used with WDS compositions determined as $x$=0.19, 0.22 and 0.25. The large slabs were cleaved into bars for four-probe resistivity measurements so that all samples were originating from the same slab of the crystal. Samples typically had a size of 2$\times$0.5$\times$0.1 mm$^3$ and long and short sides corresponding to the crystallographic $a$-axis and $c$-axis, respectively. Low-resistance contacts to the samples were made by soldering Ag wires with tin \cite{SUST,patent}. The contacts were found to be both mechanically and electronically stable under electron irradiation. Four-probe resistivity measurements were performed in a {\it Quantum Design} PPMS. Specific heat was measured in a helium cryostat by using an AC calorimeter built on SiN membrane chips at frequencies in the 1 Hz range as described in Refs.\cite{Tagliati, Willa2017}.

\begin{figure}[tb]
\begin{center}
\includegraphics[width=0.90\linewidth]{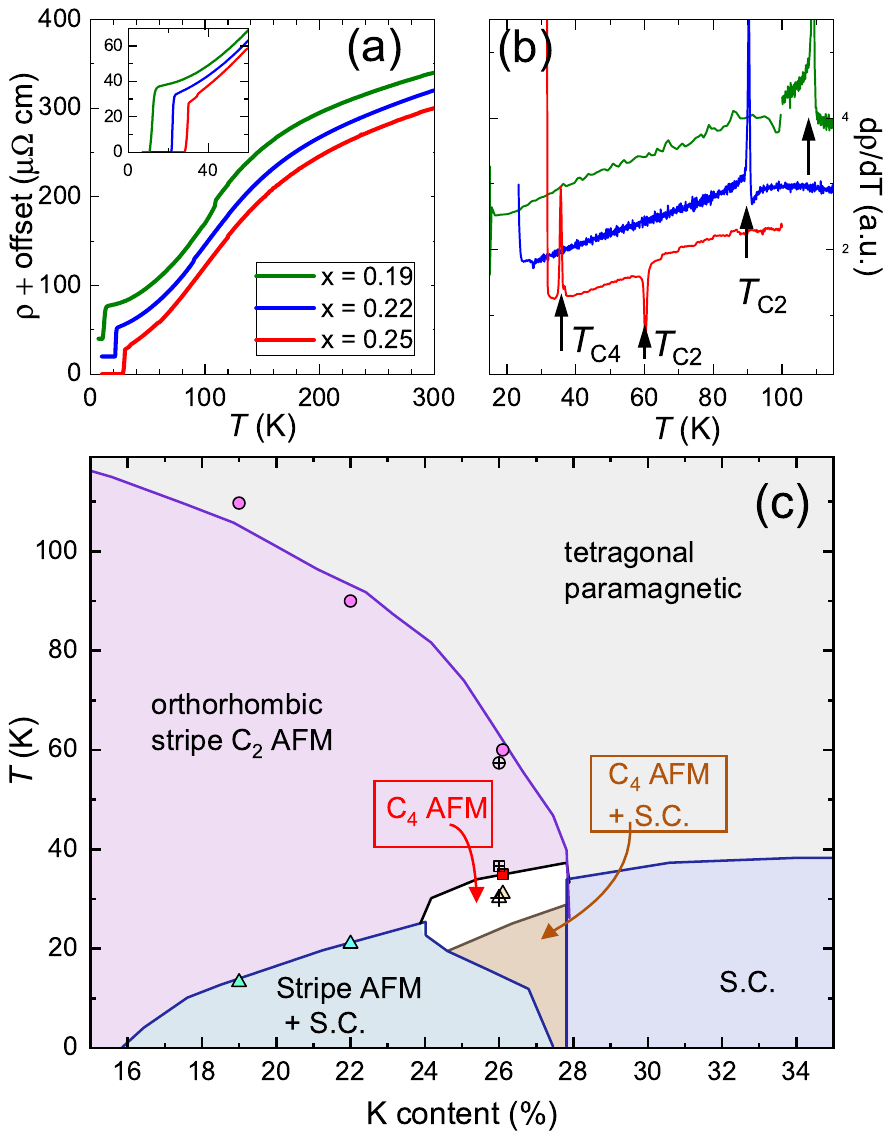}
\end{center}
\caption{
(Color Online)
(a) Temperature-dependent resistivity of selected samples of Ba$_{1-x}$K$_x$Fe$_2$As$_2$, $x=$0.19 (green), 0.22 (blue) and 0.25 (red), the curves are offset vertically.
Inset: low temperature region showing differences in the superconducting transition temperatures and resistivity values at $T_{\rm c}$.
(b) Resistivity derivative, revealing a sharp feature at the structural transitions at $T_{\rm C2}$ and $T_{\rm C4}$.
(c) Doping phase diagram of Ba$_{1-x}$K$_x$Fe$_2$As$_2$ in the range of ${\rm C_4}$ phase formation, as proposed by B\"ohmer et al. \cite{Anna} (lines). The positions of  ${\rm C_2}$ (circles), ${\rm C_4}$ (square) and superconducting (triangle) transitions for samples with $x=$0.19 and $x=$0.22 are matching well with the diagram, but the position of $x=$0.25 sample was adjusted to 0.264 to match the ${\rm C_2}$ line with the concomitant match of the ${\rm C_4}$ line.
Symbols with crosses show the positions of the features in the heat capacity measurements, see Fig.~3 below.
}
\label{resPD}
\end{figure}

For our study we selected samples with the sharpest features in the temperature-dependent resistivity $\rho(T)$ at concomitant tetragonal/orthorhombic and paramagnetic/${\rm C_2}$ antiferromagnetic transitions in samples $x=$0.19 and 0.22. The largest problem however is finding samples with sharp features at the ${\rm C_2}$ to ${\rm C_4}$ transition for $x=$0.25 which is extremely sensitive to sample to sample variation without detectable composition variations with $\Delta x \sim$0.003. We therefore did all pre-characterization of the samples with resistivity and only performed specific heat on selected samples.

The samples for resistivity measurements during and after electron irradiation were mounted on a thin mica plate in a hollow {\it Kyocera} chip, so that they could be moved between the irradiation chamber and the resistivity setup (in a different $^4$He cryostat) without disturbing the contacts.
The low-temperature 2.5 MeV electron irradiation was performed at the SIRIUS Pelletron linear accelerator operated by the \textit{Laboratoire des Solides Irradi\'{e}s} (LSI) at the \textit{Ecole Polytechnique} in Palaiseau, France \cite{SIRIUS}.
The {\it Kyocera} chip was mounted inside the irradiation chamber and was cooled by a flow of liquid hydrogen to $T \approx 22$~K in order to remove excess heat produced by relativistic electrons upon collision with the ions. The flux of electrons amounted to about 2.7 $\mu$A of electric current through a 5 mm diameter diaphragm. This current was measured with the Faraday cup placed behind a hole in the sample stage, so that only transmitted electrons were counted. The irradiation rate was about $5 \times 10^{-6}$ C$/$(cm$^{2}\cdot $s) and large doses were accumulated over the course of several irradiation runs. Throughout the manuscript we use ``pristine'' and ``unirradiated'' interchangeably to describe samples that were not exposed to electron radiation.

A selected sample $A$ of $x=$0.25 composition was irradiated multiple times adding doses in small steps and tracking the fine evolution of its temperature-dependent resistivity to determine $T_{\rm C2}$, $T_{\rm C4}$ and the superconducting $T_{\rm c}$. The sample was  extracted from the irradiation chamber following each irradiation dose step and its temperature-dependent resistivity was measured \textit{ex-situ} after annealing at room temperature. This annealing, however, does not remove residual disorder, so that the sample resistance gradually increased in successive runs. A second sample $B$ with the same composition was mounted on the same chip and underwent the same irradiation procedure. After an accumulation of a significant dose and the ensuing characterization by resistivity which produced results that were consistent with sample $A$, a small piece (100$\mu $m $ \times$ 160$\mu $m) was cut from the area between potential contacts to be used for microcalorimetric measurements. Another pristine sample was measured by specific heat as a comparison.
The samples of other compositions $x=$0.19 and $x=$0.22 were irradiated without intermediate measurements, receiving the maximum dose in one run.


\begin{figure}[tb]
\begin{center}
\includegraphics[width=0.95\linewidth]{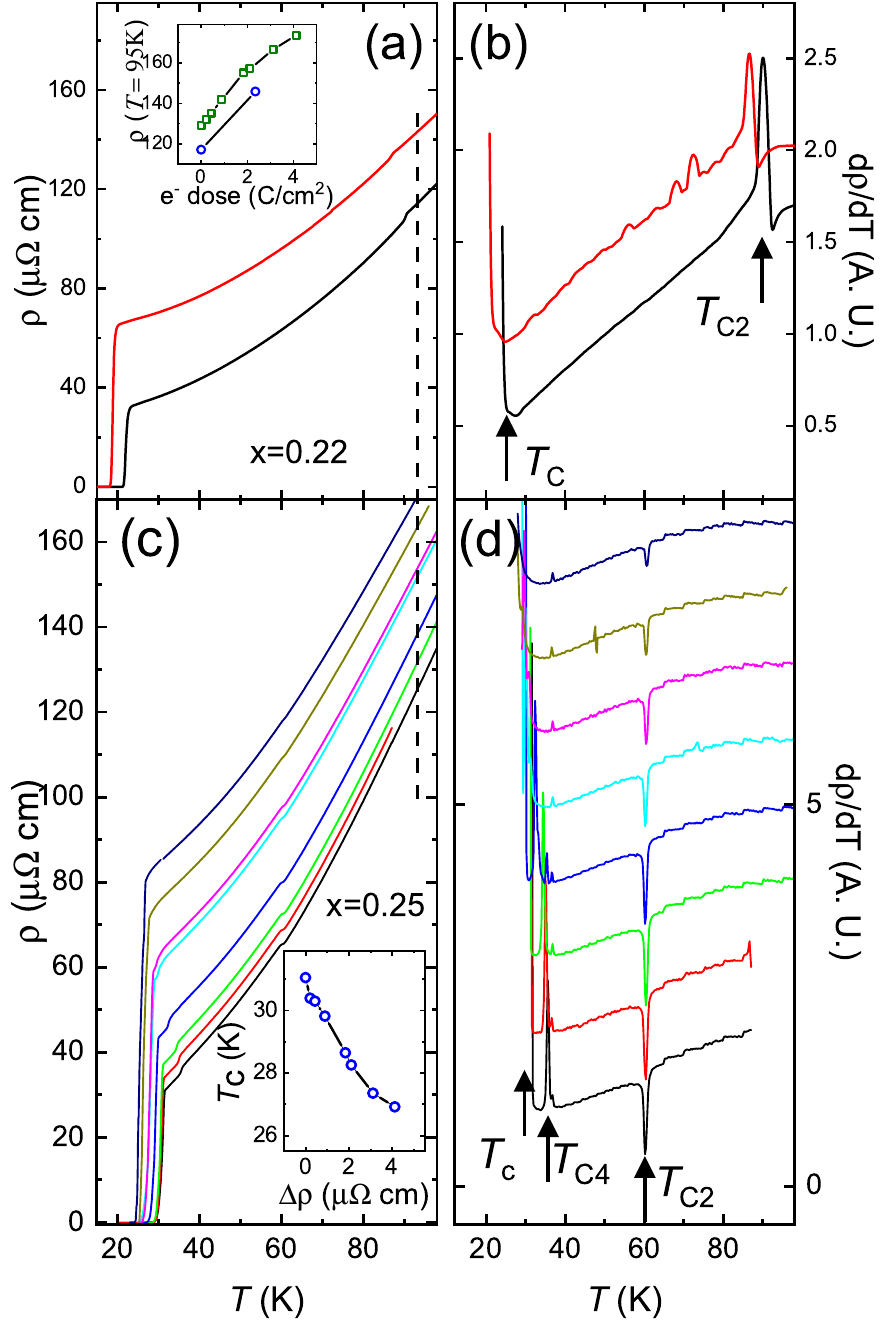}
\end{center}
\caption{(Color Online) Temperature-dependent resistivity (left panels) and resistivity derivative (right panels) of Ba$_{1-x}$K$_{x}$Fe$_2$As$_2$ with $x=$0.22 (top) and $x=$0.25 (bottom).  Black curves in panels (a) and (b) are for pristine $x=$0.22 sample, red curves are for sample after electron irradiation with 2.35 C/cm$^2$. Panels (c) and (d) show systematics of the evolution of the temperature-dependent electrical resistivity in sample with $x=$ 0.25 with irradiation, bottom to top: pristine sample (black), 0.212 C (red), 0.438 C (green), 0.893 C (blue), 1.835 C (cyan), 2.115 C (magenta), 3.115 C (dark yellow), 4.115 C (navy). Inset in panel (a) shows resistivity at 95 K as a function of electron irradiation dose for sample with $x=$0.22 (blue) and $x=$0.25 (green). Inset in panel (c) shows the evolution of the superconducting transition temperature in sample with $x=$0.25 as a function of the change of resistivity at 95~K, above $T_{\rm C2}$.
}%
\label{resder}
\end{figure}

In Fig.~\ref{resPD} we show the temperature-dependent resistivity of selected samples with $x$=0.19, 0.22 and 0.25 in the pristine state before irradiation. The room-temperature resistivity of the samples was set to 300 $\mu \Omega$cm, the statistically significant value as determined on a big array of crystals \cite{YLiucrystals}. The actually measured values for the individual samples were within the 10\% uncertainty of the geometric factor determination. The $\rho (T)$ curves show the typical behavior of hole-doped Ba$_{1-x}$K$_x$Fe$_2$As$_2$ \cite{YLiucrystals,phaseD}, with a broad crossover at around 200~K, a small acceleration of resistivity decrease on cooling through $T_{\rm C2}$ and a rather sharp superconducting transition at $T_{\rm c}$. The $T_{\rm C2}$ feature is most clearly seen as a sharp feature in the temperature derivative of the resistivity, $d\rho/dT$, top right panel of Fig.~\ref{resPD}. The resistivity of the samples just above $T_{\rm c}$ decreases monotonically with $x$ from about 40 $\mu \Omega$cm in $x=$0.19 to 30 $\mu\Omega$cm in $x=$0.25  and the residual resistivity ratios increase from about 7 to 10, respectively. The $T_{\rm C2}$ feature is shifting  down in temperature with increasing $x$ reaching $T_{\rm C2}=$60~K for $x=$0.25 (the same feature is observed at 57 K in specific heat indicating its bulk nature). In the bottom panel of Fig.~\ref{resPD} we plot the characteristic temperatures as determined from resistivity measurements (circles $T_{\rm C2}$, open up-triangles $T_{\rm c}$ as determined from offset criterion) as a function of $x$ in comparison with the phase diagram by B\"ohmer {\it et al.} \citep{Anna} (lines in the figure). The position of the $x=$0.25 sample on this phase diagram does not follow $T_{\rm C2}$ line. However, if we allow for a small variation of $x$ for our $x=$0.25 WDS sample to match $T_{\rm C2}$ with the value reported by B\"ohmer, we simultaneously match the $T_{\rm C4}$ feature (red solid square) as well. The composition difference amounts to approximately 1\%, which is presumably coming from the difference in calibrations in the composition analysis between WDS (our case) and EDX (as used by B\"ohmer {\it et al.} \citep{Anna}). The onset of the resistive transition to superconducting phase in samples A and B (not shown) occurs at 30 K with no indication of the $T_{\rm c}$ depression reported in  \cite{Anna}.

In Fig.~\ref{resder} we show the evolution of the temperature-dependent resistivity $\rho(T)$ with electron irradiation. The irradiation increases the resistivity of the samples, with the increase being nearly temperature independent above $T_{\rm C2}$, but strongly temperature-dependent below. This difference in response to controlled disorder above and below $T_{\rm C2}$ is found in other BaFe$_2$As$_2$ based materials, P-doped \cite{ShibauchiPdoped}, Ru-doped \cite{UchidaRu,Blomberg2018Ru} and K-doped \cite{Cho2014PRB,Prozorov2018}. Since the resistivity above $T_{\rm C2}$  roughly obeys Matthiessen rule, we used the post-irradiation increase of resistivity at set temperature $T$=95~K (dashed lines in left panels in Fig.~\ref{resder}) as an intrinsic measure of disorder. The electron dose-dependence of the resistivity for samples $x=$0.22 (blue circles) and $x=$0.25 (green squares) is shown in the inset in the top left panel of Fig.~\ref{resder}.

Irradiation suppresses $T_{\rm C2}$ in samples with $x=$0.19 (not shown) and $x=$0.22 (top right panel of Fig.~\ref{resder}). This is similar to the results of previous studies for all types of substitutions in BaFe$_2$As$_2$ \citep{ShibauchiPdoped,UchidaRu,Blomberg2018Ru,Cho2014PRB,Prozorov2018}. The response to irradiation in the $x=$0.25 sample is qualitatively different (bottom left panel of Fig.~\ref{resder}). While the superconducting transition temperature is monotonically suppressed with increasing resistivity, the $T_{\rm C4}$ feature moves to lower temperatures significantly faster than $T_{\rm c}$ and eventually becomes indistinguishable from the superconducting transition. Furthermore, the $T_{\rm C2}$ feature is not suppressed with increasing scattering but, in fact, a slight increase of $T_{\rm C2}$ with irradiation is found in heat capacity measurements.


\begin{figure}[tb]
\begin{center}
\includegraphics[width=0.90\linewidth]{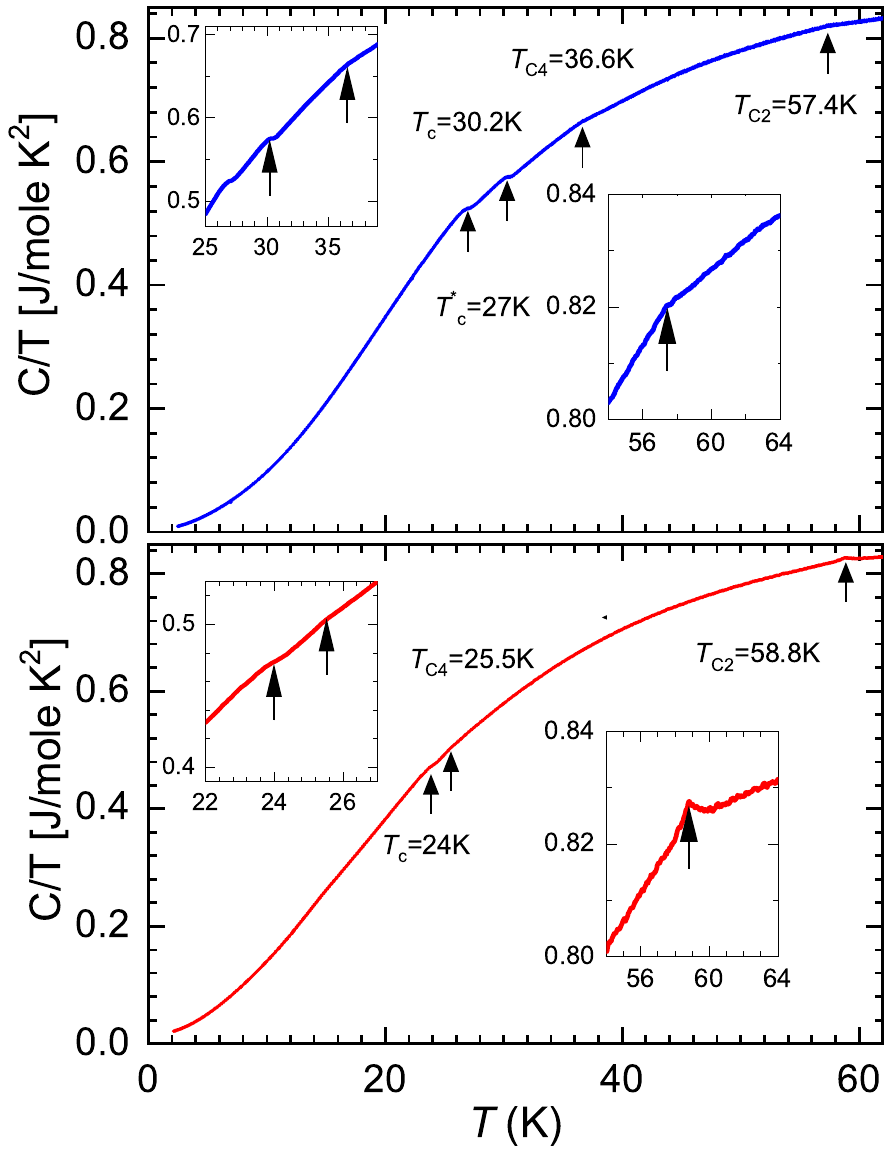}
\end{center}
\caption{(Color online) Temperature-dependent heat capacity, $C/T$, of Ba$_{1-x}$K$_x$Fe$_2$As$_2$ sample  $x=$0.25 before (top panel) and after (bottom panel) electron irradiation with 5.045~C/cm$^2$. Right insets zoom on $T_{\rm C2}$ phase transition, left insets on low-temperature transitions.
}%
\label{heatcapacity}
\end{figure}

The findings in resistivity measurements are well matched by the heat capacity measurements. In the pristine state, top panel in Fig.~\ref{heatcapacity}, clear changes of slope are seen in the $C/T$ vs $T$ plot at $T_{\rm C2}=$57.4~K, $T_{\rm C4}=$36.6~K as well as two low-temperature features corresponding to the superconducting transition and possibly the reentrant ${\rm C_2}$ phase. These features are shown with crossed symbols in Fig.~\ref{resPD} above. $T_{\rm C4}$ is strongly suppressed after irradiation, faster than the superconducting transition, while the ${\rm C_2}$ transition becomes sharper and moves slightly up in temperature.

\begin{figure}[tb]
\begin{center}
\includegraphics[width=0.90\linewidth]{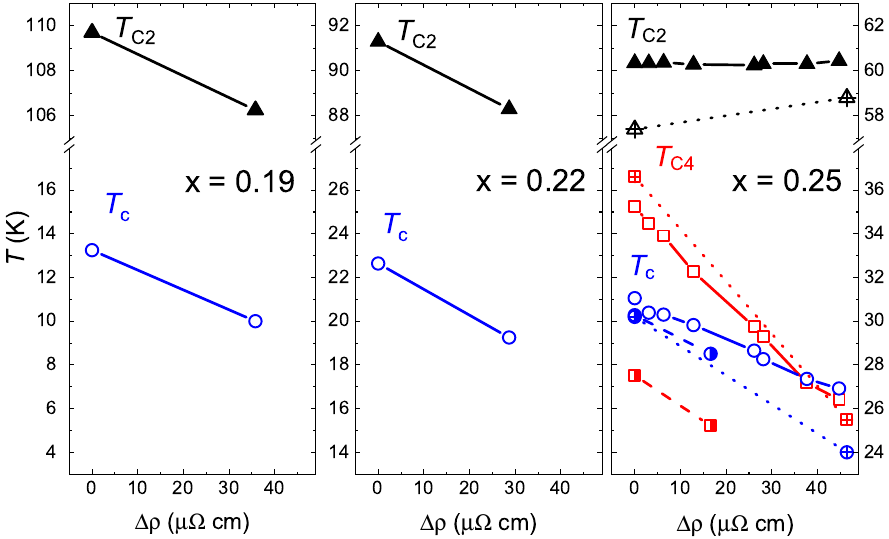}
\end{center}
\caption{(Color online) Transition temperatures of samples of Ba$_{1-x}$K$_x$Fe$_2$As$_2$ x=0.19 (left panel), x=0.22 (center panel), and x=0.25 (right panel)  as a function of scattering rate increase characterized with resistivity increase in the paramagnetic tetragonal phase above $T_{\rm C2}$. Note the similar rates of superconducting $T_{\rm c}$ suppression in all compositions, the fast suppression of ${\rm C_2}$ phase in samples $x=$0.19 and 0.22, the two times faster suppression of ${\rm C_4}$ phase in $x=$0.25 and the slight increase of $T_{\rm C2}$ with irradiation in $x=$0.25.
}%
\label{PD}
\end{figure}

In Fig.~\ref{PD} we summarize our observations as plots of characteristic temperatures for Ba$_{1-x}$K$_x$Fe$_2$As$_2$ as a function of change of resistivity after irradiation. Left panel is data for the sample with $x=$0.19, middle panel for $x=$0.22 and right panel is for $x=$0.25.
Note that the rates of the superconducting transition suppression with disorder, 0.091, 0.118 and 0.098 $K/\mu \Omega$cm for $x$=0.19, 0.22 and 0.25 respectively,  are very close to each other and to the rate of the $T_{\rm C2}$ suppression, 0.096, and 0.105 for $x=$0.19 and $x=$0.22. The rate of the ${\rm C_4}$ phase suppression in sample $x=$0.25, 0.21 $K/\mu \Omega{\rm cm}$ in resistivity and 0.24 $K/\mu \Omega{\rm cm}$ in heat capacity measurements, is about two times faster than that of ${\rm C_2}$ phase suppression in $x=$0.19 and $x=$0.22 samples. This rate is also significantly higher than the rate of ${\rm C_4}$ spin-vortex phase suppression in CaK(Fe$_{1-x}$Ni$_x$)$_4$As$_4$, 0.16 $K/\mu \Omega{\rm cm}$ \cite{Serafim}. A slight increase of $T_{\rm C2}$ in sample $x=$0.25 is found after irradiation in heat capacity measurements.

It is most natural to explain our findings as evidence for competition between the ${\rm C_2}$ and the ${\rm C_4}$ phases, with a suppression of the ${\rm C_4}$ phase leading to a stabilization of the ${\rm C_2}$ phase. Interestingly, this behavior is found for a certain parameter range in the calculations of Hoyer {\it et al.} \cite{disorderSchmalian,note}, though this paper considers the case of phase competition near the magnetic transition temperature as opposed to the case of the ${\rm C_4}$ phase existing deep in the domain of the ${\rm C_2}$ phase as found in our experiment.


In conclusion, we find that controlled disorder introduced by low-temperature irradiation with relativistic 2.5 MeV electrons rapidly suppresses the transition temperature between antiferromagnetic ${\rm C_2}$ and ${\rm C_4}$ phases and leads to the relative stabilization of the ${\rm C_2}$ phase. This behavior can be found for the  parameter range characterized by weak nesting in the itinerant electron magnetism model by Hoyer {\it et al.} \cite{disorderSchmalian}, though the phase stability relations were considered only at the transition temperature for magnetic ordering. Our findings suggest that further theoretical analysis that will consider possible 1st order transition between two phases, hence, phase coexistence and possible separation, may be necessary.

We thank R.~M.~Fernandes, P.~P.~Orth and J.~Schmalian for fruitful discussions. We thank B. Boizot and all SIRIUS team for operation of electron irradiation at SIRIUS facility at {\'E}cole Polytechnique, Palaiseau, France. This work was supported by the US Department of Energy, Office of Science, BES Materials Science and Engineering Division. Irradiation experiments were supported by the EMIR (R\'eseau national d’acc\'el\'erateurs pour les \'Etudes des Mat\'eriaux sous Irradiation) network, proposal 16-2125. K.W. acknowledges support from the Swiss National Science Foundation through the Postdoc Mobility program.



\begin{thebibliography}{99}

\bibitem{Mathur}
N. D. Mathur, F.M. Grosche, S. R. Julian, I. R. Walker, D. M. Freye, R. K.W. Haselwimmer, and G. G. Lonzarich,
Nature (London) {\bf 394}, 39 (1998).

\bibitem{heavyfermionLonzarich}
B.~D.~White, J.~D.~Thompson, and M.~B.~Maple, Physica C- Supercond. and its applications,
{\bf 514}, 246 (2015).



\bibitem{LouisNatureQCP}
B.~Keimer, S.~A.~Kivelson, M.~R.~Norman, S.~Uchida, and J.~Zaanen,
Nature(London) {\bf 518}, 179 (2015).

\bibitem{Paglione}
Johnpierre~Paglione, and Richard~L.~Greene,
Nature Phys. {\bf 6}, 645 (2010).


\bibitem{review}
P.~Monthoux, D.~Pines, and G.~G.~Lonzarich,
Nature (London) {\bf 450}, 1177 (2007).

\bibitem{KchoNbSe2} K.~Kyuil Cho, M. Konczykowski, S. Teknowijoyo, M. A. Tanatar, J. Guss, P. B. Gartin, J. M. Wilde, A. Kreyssig, R. J. McQueeney, A. I. Goldman, V. Mishra, P. J. Hirschfeld, and R. Prozorov, Nature Comm. {\bf 9}, 2796 (2018).

\bibitem{PdopedQCP} Y. Nakai, T. Iye, S. Kitagawa, K. Ishida, H. Ikeda, S. Kasahara, H. Shishido, T. Shibauchi, Y. Matsuda, and T. Terashima,
Phys. Rev. Lett. {\bf 105}, 107003 (2010).

\bibitem{Pdopedcaxis}
M. A. Tanatar, K. Hashimoto, S. Kasahara, T. Shibauchi, Y. Matsuda, and R. Prozorov, Phys. Rev. B {\bf 87}, 104506 (2013).



\bibitem{PAnalytis} Ian~M.~Hayes, Ross~D.~McDonald, Nicolas~P.~Breznay, Toni~Helm, Philip~J.~W.~Moll, Mark~Wartenbe, Arkady~Shekhter, James~G.~Analytis,
Nature Phys. {\bf 12}, 916 (2016).

\bibitem{stripe}
Clarina de la Cruz, Q. Huang, J. W. Lynn, Jiying Li, W. Ratcliff II, J. L. Zarestky, H. A. Mook, G. F. Chen, J. L. Luo, N. L. Wang, and Pengcheng Dai,
Nature (London) {\bf 453}, 899 (2008).


\bibitem{stripe2}
Marianne Rotter, Marcus Tegel, Dirk Johrendt, Inga Schellenberg, Wilfried Hermes, and Rainer Pöttgen
Phys. Rev. B {\bf 78}, 020503(R) (2008).




\bibitem{Elena}
E. Hassinger, G. Gredat, F. Valade, S. R. de Cotret, A. Juneau-Fecteau, J.-Ph. Reid, H. Kim, M. A. Tanatar, R. Prozorov, B. Shen, H.-H. Wen, N. Doiron-Leyraud, and Louis Taillefer
Phys. Rev. B {\bf 86}, 140502(R) (2012).

\bibitem{Elena2}
E. Hassinger, G. Gredat, F. Valade, S. René de Cotret, O. Cyr-Choinière, A. Juneau-Fecteau, J.-Ph. Reid, H. Kim, M. A. Tanatar, R. Prozorov, B. Shen, H.-H. Wen, N. Doiron-Leyraud, and Louis Taillefer
Phys. Rev. B {\bf 93}, 144401 (2016).


\bibitem{Anna}
A. E. B\"ohmer, F.~Hardy, L.~Wang, T.~Wolf, P.Schweiss, and C.~Meingast, Nat. Comm. {\bf 6},  7911 (2015).

\bibitem{AvciNa}
S. Avci, J. M. Allred, O. Chmaissem, D. Y. Chung, S. Rosenkranz, J. A. Schlueter, H. Claus, A. Daoud-Aladine, D. D. Khalyavin, P. Manuel, A. Llobet, M. R. Suchomel, M. G. Kanatzidis, and R. Osborn,
Phys. Rev. B {\bf 88}, 094510 (2013).



\bibitem{AvciK}
J. M. Allred, S. Avci, D. Y. Chung, H. Claus, D. D. Khalyavin, P. Manuel, K. M. Taddei, M. G. Kanatzidis, S. Rosenkranz, R. Osborn, and O. Chmaissem, Phys. Rev. B {\bf 92}, 094515 (2015).

\bibitem{entropy}
L. Wang, F. Hardy, A. E. B\"ohmer, T. Wolf, P. Schweiss, and C. Meingast,
Phys. Rev. B {\bf 93}, 014514 (2016).


\bibitem{Furukawa}
Q.-P. Ding, W. R. Meier, J. Cui, M. Xu, A. E. B\"ohmer, S. L. Bud'ko, P. C. Canfield, and Y. Furukawa,
Phys. Rev. Lett. {\bf 121}, 137204 (2018).

\bibitem{CaNa}
K. M. Taddei, J. M. Allred, D. E. Bugaris, S. H. Lapidus, M. J. Krogstad, H. Claus, D. Y. Chung, M. G. Kanatzidis, R. Osborn, S. Rosenkranz, and O. Chmaissem,
Phys. Rev. B {\bf 95}, 064508 (2017).


\bibitem{SrNa}
K. M. Taddei, J. M. Allred, D. E. Bugaris, S. Lapidus, M. J. Krogstad, R. Stadel, H. Claus, D. Y. Chung, M. G. Kanatzidis, S. Rosenkranz, R. Osborn, and O. Chmaissem,
Phys. Rev. B {\bf 93}, 134510 (2016)

\bibitem{Mn} M. G. Kim, A. Kreyssig, A. Thaler, D. K. Pratt, W. Tian, J. L. Zarestky, M. A. Green, S. L. Bud’ko, P. C. Canfield, R. J. McQueeney, and A. I. Goldman,
Phys. Rev. B {\bf 82}, 220503(R) (2010).




\bibitem{doubleQ} J.~M.~Allred, K.~M.~Taddei, D.~E.~Bugaris, M.~J.~Krogstad, S.~H.~Lapidus, D.~Y.~Chung, H.~Claus, M.~G.~Kanatzidis,
D.~E.~Brown, J.~Kang, R.~M.~Fernandes, I.~Eremin, S.~Rosenkranz, O.~Chmaissem, and R.~Osborn,
Nature Phys. {\bf 12}, 493 (2016).

\bibitem{hedgehog}
William~R.~Meier,  Qing-Ping Ding, Andreas Kreyssig, Sergey~L.~Bud'ko, Aashish~Sapkota, Karunakar~Kothapalli, Vladislav~Borisov, Roser	Valenti,
 Cristian~D.~Batista, Peter~P.~Orth, Rafael~M.~Fernandes, Alan~I.~Goldman, Yuji Furukawa,  Anna~E.~B\"ohmer, Paul C.~Canfield,
NPJ Quantum Materials {\bf 3}, 5 (2018).

\bibitem{RMF} R. M. Fernandes, S. A. Kivelson, and E. Berg,
Phys. Rev. B {\bf 93}, 014511 (2016).

\bibitem{Efremov}
Jing Wang, Guo-Zhu Liu, Dmitry V. Efremov, and Jeroen van den Brink,
Phys. Rev. B {\bf 95}, 024511 (2017).

\bibitem{QSi} Rong Yu, Ming Yi, Benjamin A. Frandsen, Robert J. Birgeneau, Qimiao Si, arxiv.1706.07087

\bibitem{Orth1}
Morten H. Christensen, Peter P. Orth, Brian M. Andersen, and Rafael M. Fernandes, Phys. Rev. B {\bf 98}, 014523 (2018).


\bibitem{Orth2}
Morten H. Christensen, Peter P. Orth, Brian M. Andersen, and Rafael M. Fernandes,
Phys. Rev. Lett. {\bf 121}, 057001 (2018).

\bibitem{disorderSchmalian}
Mareike Hoyer, Rafael M. Fernandes, Alex Levchenko, and J\"org Schmalian,
Phys. Rev. B {\bf 93}, 144414 (2016).

\bibitem{Serafim} S. Teknowijoyo, K. Cho, M. Kończykowski, E. I. Timmons, M. A. Tanatar, W. R. Meier, M. Xu, S. L. Bud'ko, P. C. Canfield, and R. Prozorov,
Phys. Rev. B {\bf 97}, 140508(R) (2018).


\bibitem{YLiucrystals}
Y. Liu, M. A. Tanatar, W. E. Straszheim, B. Jensen, K. W. Dennis, R. W. McCallum, V. G. Kogan, R. Prozorov, and T. A. Lograsso, Phys. Rev. B {\bf 89}, 134504 (2014).

\bibitem{SUST} M. A. Tanatar, N. Ni, S. L. Bud'ko, P. C. Canfield, and R. Prozorov,  Supercond. Sci. Technol. {\bf 23}, 054002 (2010).

\bibitem{patent} M.~A.Tanatar, R.~Prozorov, N.~Ni, S. L. Bud'ko, P. C. Canfield, U.S. Patent 8,450,246 (Sept.1, 2011).

\bibitem{Tagliati}
S. Tagliati, V. M. Krasnov, and A. Rydh, Rev. Sci. Instrum. {\bf 83}, 055107 (2012).

\bibitem{Willa2017}
K. Willa, S. Diao, D. Campanini, U. Welp, R. Divan, M. Hudl, Z. Islam, W.-K. Kwok, and A. Rydh, Rev. Sci. Instrum. {\bf 88}, 125108 (2017).



\bibitem{SIRIUS} http://emir.in2p3.fr/ LSI, electron irradiation facility.


\bibitem{phaseD} M. A. Tanatar, W. E. Straszheim, Hyunsoo Kim, J. Murphy, N. Spyrison, E. C. Blomberg, K. Cho, J.-Ph. Reid, Bing Shen, Louis Taillefer, Hai-Hu Wen, and R. Prozorov,
Phys. Rev. B {\bf 89}, 144514 (2014).


\bibitem{ShibauchiPdoped}
Yuta Mizukami, Marcin Konczykowski, Kohei Matsuura, Tatsuya Watashige, Shigeru Kasahara, Yuji Matsuda, and Takasada Shibauchi,
J. Phys. Soc. Jpn {\bf 86}, 083706 (2017).

\bibitem{UchidaRu} L. Liu, T. Mikami, S. Ishida, K. Koshiishi, K. Okazaki, T. Yoshida, H. Suzuki, M. Horio, L. C. C. Ambolode, II, J. Xu, H. Kumigashira, K. Ono, M. Nakajima, K. Kihou, C. H. Lee, A. Iyo, H. Eisaki, T. Kakeshita, S. Uchida, and A. Fujimori,
Phys. Rev. B {\b 92}, 094503 (2015).

\bibitem{Blomberg2018Ru}
E.~C.~Blomberg, M.~A.~Tanatar, A.~Thaler, S. L. Bud'ko, P. C. Canfield, and R. Prozorov,  J. Phys. Cond. Matt. {\bf  30}, 315601 (2018).

\bibitem{Cho2014PRB} K. Cho, M. Ko\'nczykowski, J. Murphy, H. Kim, M. A. Tanatar, W. E. Straszheim, B. Shen, H. H. Wen, and
R. Prozorov,  Phys. Rev. B {\bf 90}, 104514 (2014).


\bibitem{Prozorov2018}
R. Prozorov, M. Konczykowski, M. A. Tanatar, H. H. Wen, R. M. Fernandes, P. C. Canfield,  arXiv:1808.09532

\bibitem{note} This can be seen most clearly as a slight expansion of the (green, SM) domain of ${\rm C_2}$ phase in the central area of Fig.~2 in the paper by Hoyer {\it et al.}, \cite{disorderSchmalian}.








\end{thebibliography}
\end{document}